# Evaluation of methods for the determination of tortuosity of Li-ion battery separators

Wei Sun [a], Q.M.Li [a*], Ping Xiao [b], Paola Carbone [c]

a. Department of Mechanical, Aerospace and Civil Engineering, School of Engineering, The University of Manchester, Manchester M13 9PL, UK

b. Department of Materials, Henry Royce Institute, The University of Manchester, Manchester M13 9PL, UK

c. Department of Chemical Engineering, School of Engineering, The University of Manchester, Manchester M13 9PL, UK

 * corresponding author; e-mail: Qingming.li@manchester.ac.uk

**Abstract:** The porosities and tortuosities are commonly utilized to characterize the microstructure of a Li-ion battery's separator and are adopted as key input parameters in advanced battery models. Herein, a general classification of the tortuosity for a porous medium is introduced based on its bi-fold significance, i.e., the geometrical and physical tortuosities. Then, three different methods for the determination of separator's electrical tortuosity are introduced and compared, which include the empirical Bruggeman equation, the experimental method using Electrochemical Impedance Spectrum (EIS) testing and the numerical method using realistic 3D microstructure of the separator obtained from nanoscale X-ray Computed Tomography (XCT). In addition, the connection between the geometrical tortuosity and the electrical tortuosity of a separator is established by introducing the electrical phenomenological factor ($\beta_e$), which can facilitate the understanding of the relationship between the microstructure characteristics and transport properties of the separators. Furthermore, to quantitively compare the values of the tortuosities determined by different methods, the corresponding effective transport coefficients ($\delta$) are compared, which was usually used as a correction for effective diffusivity and conductivity of electrolytes in porous media.

**Keywords:** Li-ion battery separator; tortuosity; porosity; 3D microstructure; XCT; resistivity.

# 1 Introduction

Li-ion batteries (LIBs) have been widely adopted as energy storage devices in various applications due to their high energy density and good electrochemical performance. A basic LIB cell consists of two porous electrodes coated on metallic current collectors and a porous polymeric separator [1]. The porous separator filled with electrolyte allows the Li-ion to move through between the electrodes while preventing an internal short circuit from direct contact between the positive and negative electrodes. To accurately evaluate the safety, performance and degradation of LIB at multiple scales, it is necessary to ensure that the ion transport through battery's separator can be reliably described in a battery model.

Homogenised multi-physics battery models have been widely adopted to predict battery performance with high efficiency [2-7], among which the model proposed by Newman [3] is the most popular and has been implemented in COMOSL Multiphysics software where the porous electrodes and separator are homogenized using the macroscopic description of porous media. The homogenized battery model considers the effect of the separator's microstructure on the macroscopic performance of LIB using two important characterization parameters of microstructure, i.e., porosity ($\Phi$) and tortuosity ($\tau$), to estimate the effective diffusivity ($D_{eff}$) and effective conductivity ($\sigma_{eff}$) of the separator immersed in the electrolyte [7], i.e.

$$D_{eff} = \frac{\Phi}{\tau^2} \cdot D_0 \tag{1}$$

$$\sigma_{eff} = \frac{\Phi}{\tau^2} \cdot \sigma_0 \tag{2}$$

where $D_0$ and $\sigma_0$ are respectively the bulk electrolyte diffusivity and bulk electrolyte conductivity.

Tortuosity plays an important role in mass transfer, charge transfer and ohmic loss in a LIB [1, 8]. The tortuosity ($\tau_{Brugg}$) of a separators was empirically related to its porosity via Bruggeman equation [9], i.e.

$$\tau_{Brugg}^2 = \Phi^{1-\alpha} \qquad (3)$$

where $\alpha$ is the Bruggeman exponent, which is assumed to be 1.5 in its standard form. Because it is much easier to measure the porosity of a separator than to measure its tortuosity, Eq. (3) has been commonly used to determine the tortuosity of a separator from its porosity as a key input parameter of homogenised LIB model to predict LIB's electrochemical performance [3, 7, 10-12]. However, Eq.(3) was proposed based on experiments on porous materials with spherical bead as the solid phase, which is quite different from the complex pore geometry of a separator. Therefore, Eq.(3) should be further evaluated for its applicability to separators.

Two experimental techniques have been developed for the determination of tortuosities of various separators, i.e. the impedance-based techniques to measure the conductivity or resistivity of a separator immersed with electrolyte [13-21] and the polarization-interrupt method to measure the diffusivity of a separator immersed with electrolyte [16]. Limited experimental evidences have demonstrated that the microstructure of a separator has a great influence on the capacity performance of LIB [13, 15, 19, 22]. For example, it has been shown in [19] that the capacity performances of LIB with separator Celgard 2500 under 2 and 3 C-rates are respectively 57% and 47% higher than the capacity performances of LIB with separator Celgard 2325. The reported porosities of Celgard 2500 and Celgard 2325 are respectively 55% and 39% while their tortuosities are 1.70 and 1.98, respectively. Furthermore, four different types of separators were tested in [22] indicating that the assembled LIBs with separators of higher porosity have relatively lower battery capacities, which can be attributed to the initial loss of lithium and the decomposition of the electrolyte to form interphase layers. More interestingly, it was noted by the present authors that the reported values of tortuosity for the same separator vary quite significantly. For example, the reported tortuosities for the separator of Celgard 2500 are 3.2 [13], 2.16 [14], 2.5 [15], 1.58 [18] and 1.7 [19], respectively. These large discrepancies imply a large inconsistency in the determination of the separator's tortuosity, which needs to be resolved due to its important role in the homogenised battery model. In addition, due to the lack of experimental details, such as cell setup, the thickness of the separator, electrode area, electrolyte selection and the ambient temperature during testing, the accuracy of the reported tortuosities is questionable and should be carefully assessed. Although various experimental setups to determine a separator's tortuosity have been reported, e.g. a coin-cell format used in [14, 15, 19, 20], a pouch-cell format in [16-18, 21] and a self-designed copper block setup in [17, 18], there is a lack of comparison among these different methods for the determinations of a separator's tortuosity. Therefore, it is necessary to further study the method for the determination of a separator's tortuosity.

With the rapid development of imaging technology, researchers started to determine the tortuosity of a separator numerically using the reconstruction of the 3D microstructure of the separator from nanoscale X-ray computed tomography (XCT) [23, 24], combined with focused ion beam and scanning electron microscope (FIB-SEM) [8, 25], or both the nanoscale XCT and SEM [26]. It is noted that the porosity of the separator Celgard 2500 (53.3%) obtained from nanoscale XCT is lower than the value of the manufacturer's specification (55%) [23, 24], which is counterintuitive because the porosity measured from XCT should be larger than the manufacturer's porosity obtained from porosity tests since some fine fibrils of the separator cannot be captured by nanoscale XCT due to its intrinsic resolution limitation. Xu, et al. [26] tried to extend the above limitation by reconstructing the microstructure of the separator by adding fibrils via the stochastic method based on the XCT dataset and the SEM surface observation. It was found in [17] that the transport properties of the separator are sensitive to the number of fibrils added to the microstructure of the separator, which will have a strong impact on the fast-charging model of LIB. Unfortunately, the right number of the added fibrils cannot be determined in their study due to the lack of reliable reference values of the tortuosity, e.g., the tortuosity values determined by reliable experimental methods. As a result, it is worthwhile to establish a connection between the microstructural geometry and the macroscopic performance of the porous separator. Furthermore, due to the thin-thickness nature of the separator in the through-thickness direction (TTD) (i.e., through-plane direction), the existing experimental method can only be used to determine the tortuosity of a separator along the TTD. However, the tortuosities along both the TTD and the in-plane direction of a separator can be determined by the numerical method using the separator's 3D microstructure. Therefore, it is necessary to examine the correlation between the tortuosity determined from the experimental method and the tortuosity determined from the microstructure of the separator.

To fill the above gaps and improve the understanding of the role of tortuosity in the function of a separator, three independent methods for determining the tortuosity of the dry-processed separator Celgard 2500 are introduced and compared. Firstly, the definitions of tortuosity in the field of the porous separator are discussed with considering its bi-fold significance. Then, the empirical Bruggeman equation based on the porosity-tortuosity relationship was used to determine the tortuosity of the separator. Subsequently, the tortuosity of the separator was experimentally determined using impedance testing, from which highly consistent impedance results were achieved. Furthermore, the tortuosities along TTD and in-plane directions were determined based on the reconstruction of the 3D microstructure obtained from nanoscale XCT. To quantitively compare those tortuosities determined by different methods, the effective transport coefficient is introduced in this study. Finally, based on the obtained tortuosities from impedance testing and the simulation method using the XCT image along the TTD, the tortuosities of the separator along in-plane directions are analytically predicted.

## 2 Tortuosity and its measurement methods

Since multiple definitions of tortuosity and various determination methods have been introduced in different literatures, which may confuse the readers, it is necessary to clarify the definitions of tortuosity before describing the methodology used in this study. Generally, tortuosity is a terminology to describe the sinuosity and interconnectedness of the pore space in a porous medium, which is a key factor to control the transport process through porous media. Although tortuosity has its unique geometrical significance, its measurement is usually associated with a particular physical process, i.e. it has bi-fold significance. It implies that the macroscopic physical behaviour of a porous medium is intrinsically determined by its geometrical microstructure and the occurrence of the physical process in the microstructure. In the literature, tortuosity has been defined as either a geometrical parameter or one that is related to a particular physical mechanism, e.g. hydraulic, electrical or diffusive mechanism [27]. Therefore, tortuosity is both a geometrical characteristic and a physical characteristic of the porous medium.

### 2.1 Geometrical tortuosity and determination method

Geometrical tortuosity is defined as the ratio of the effective length of the flow path to the straight-line length ($L$) in the macroscopic flow direction [27, 28]. As illustrated in Fig. 1, the transport flow of the porous medium is shaped by the channel of the pore, and thus, the shortest flow path ($L_g$) is frequently adopted as the effective flow path ($L_h$) to determine the geometrical tortuosity $\tau_g$ and can be expressed as [29]

$$\tau_g = \frac{L_h}{L} = \frac{L_g}{L} \qquad (4)$$

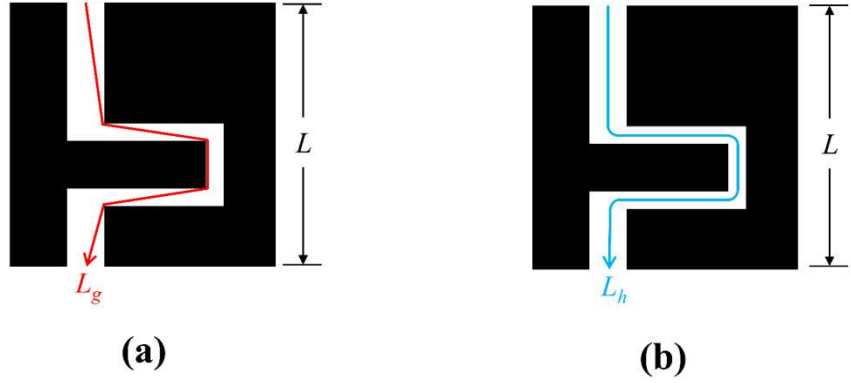

Fig. 1 (a) The shortest flow path $L_g$, (b) the effective flow path $L_h$ and the straight-line length in the 2D view.

Hence, the geometrical tortuosity can be regarded as the microstructural characterization of the sinuosity and interconnectedness of a porous medium, which is determined purely by its microstructures. Various image-based analysis methods and associated algorithms have been used to determine the geometrical tortuosity from the reconstructed 2D/3D microstructure of the porous media, such as the direct shortest method (DSPSM) [30], the skeleton shortest-path search method (SSPSM) [31], the fast-marching method (FMM) [32, 33] and the pore centroid method (PCM) [34, 35], etc. All these methods are directly operated based on the pixel/voxel image data whose implementations are straightforward and computationally efficient.

## 2.2 Physical tortuosities

Compared to the geometrical tortuosity characterising the microstructure of the porous medium, the physical tortuosities are related to a specific macroscopic physical transport process or phenomenon, such as the transport of mass, charge, energy or other physical quantities. Therefore, depending on the actual physical transport phenomenon in the porous medium, various physics-based tortuosities, which are termed as physical tortuosities in the present study, were introduced for different physical transport mechanisms, e.g., the hydraulic, electrical, diffusional and thermal tortuosities. Herein, since our interest is in the ionic transport in the porous structure of the separator, we only focus on the electrical tortuosity. However, the methodology developed in this study is generally applicable to other physical tortuosities. Detailed and systematic reviews for various physical tortuosities can be found in [27, 28, 36].

The electrical tortuosity ($\tau_e$) is determined through the electrical conductivity experiment, in which the effective conductivity ($\sigma_{eff}$) of the porous medium sample saturated with an electrolyte is measured. Due to the retarding effect of the porous structure on the transport of charges, the measured effective conductivity is smaller than the bulk electrolyte conductivity ($\sigma_0$) [37], and their relationship can be expressed as [13, 38]

$$\sigma_{eff} = \sigma_0 \frac{\Phi}{\tau_e^2} \tag{5}$$

which will be further discussed in Section 3.2.3.

## 3 Material descriptions and methodology

### 3.1 Separators and electrolyte

The commercial separator Celgard 2500 made from polypropylene (PP) through a dry-processed uniaxial stretching in the machine direction (MD) is used in this study. A more detailed manufacturing process for Celgard 2500 can be found in [39]. This porous separator has a monolayer structure with a nominal thickness of 25 μm. The properties of the Celgard 2500 given by the manufacturer are shown in Table 1.

Table 1 The properties of Celgard 2500 given by the Manufacturer [23]

| Sample | Thickness | Porosity (%) | Average pore diameter (µm) | Gurley number (s) |
|---|---|---|---|---|
| Celgard 2500 | 25 µm | 55 | 0.064 | 200 |

The electrolyte solution of 1M LiPF$_6$ in a 3:7 (v:v) mixture of ethylene carbonate (EC) and ethyl methyl carbonate (EMC) was purchased from 'dodochem.net' and used in the impedance measurement. According to the recommendation in [18], it is necessary to consider the conductivity of the bulk electrolyte in the electrical conductivity experiment where the bulk electrolyte with conductivity in the range of 0.3-1.0 S/m offers the best measurement condition. The bulk electrolyte conductivity used in our experiment is 0.830 S/m taken from [40]. Therefore, the adopted electrolyte is appropriate for the determination of the electrical tortuosity of the separator.

### 3.2  Determination of the electrical tortuosity using the experimental method

#### 3.2.1  Thickness and porosity measurement

To determine the tortuosity of a separator, the thickness and porosity of the separator need to be measured experimentally. To accurately measure the thickness of the separator, a digital micrometre with a precision of 1 µm was used according to the standard procedure (TAPPI T441 method). To ensure the accuracy of the thickness measurement, the top ratchet of the micrometre will stop when the contact force reaches 5-10 N. The measurement was repeated 3 times and the thickness of the separator is 25.06 ± 0.05 µm, which was obtained using the averaged value from 40 layers of the separators stacked together.

The porosity was measured at an ambient temperature of 21 °C according to the following procedure, i.e. (i) The separator is punched into a circular disk with a diameter of 20 mm; (ii) The weight of the separator disk ($w_1$) is measured in the air environment using a semi-micro analytical balance (Pioneer PX5 from OHAUS) with an accuracy of 0.00001 g; (iii) The separator disk is soaked in the absolute ethanol and its weight ($w_2$) is measured again; (iv) The porosity ($\Phi$) of the separator is obtained according to Archimedes' principle using the following equation [41]

$$\Phi = 1 - (\theta \frac{w_1 - w_2}{\rho_1 - \rho_2})/V_s \qquad (6)$$

where $\theta$ is the balance correction factor (0.99985) when the air buoyancy of the adjustment weight is considered; $\rho_1 = 0.7886\ g/cm^3$ is the density of absolute ethanol at 21°C ; $\rho_2 = 0.0012\ g/cm^3$ is the air density; $V_s$ is the macroscopic volume of the single-layer separator.

### 3.2.2 Coin cell fabrication and impedance measurements

As described in Eq. (5), to determine the electrical tortuosity of the separator Celgard 2500 along the TTD, it is necessary to measure the effective conductivity of the separator immersed with electrolyte. Therefore, as shown in Fig 2 (a), the configured coin cell with separators was selected to measure the ionic resistance aroused by the separators. In addition, to quantify the resistance aroused by the component of the coin cell, an empty coin cell without a separator (but with identical rest parts) was adopted and illustrated in Fig 2. (b). The detailed experimental procedures are described below.

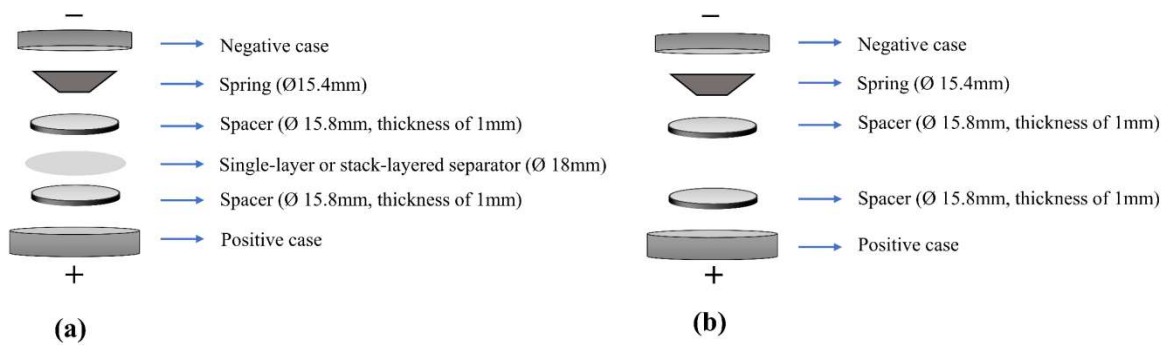

Fig. 2 (a) Designed coin-cell structure with added separator, and (b) empty cell without added separator for impedance measurement.

The single-layered or stack-layered separators (25 μm to 100 μm) were punched to produce a circular disk with an 18 mm diameter. It is worthwhile to note that the stack-layered separators need to be folded together before they are punched in order to achieve repeatable test results. Subsequently, the single-layered or stack-layered separators wetted with an electrolyte solution of 1M $LiPF_6$ in EC/EMC (3:7 v:v) were placed inside the CR2032 coin cell between the two stainless steel spacers with a nominal thickness of 1mm. The spring was placed between the stainless-steel spacer and the negative case. The coin-cell cases, stainless steel spacers, and springs were obtained from MTI corporation. The whole fabrication process of the coin cell was done in the argon-filled environment inside Pure Lab HE Glovebox with $O_2 \leq 0.1$ and $H_2O$ level = 0.0 ppm. During the assembly process, the coin cells were crimped using MSK-110 in the pressure range of 500-700 psi (~ 3.48 to 4.83 MPa). Since the separator of Celgard 2500 only has small compressive deformation when subjected to this level of pressure, as shown in our previous compression tests on this separator [24], the induced compressive deformation can be reasonably neglected. Three identical coin cells were assembled for each thickness of the separator to achieve consistency. The impedances of the assembled coin-cells and the empty cells were then evaluated by electrochemical impedance spectra (EIS) test using electrochemical workstation CS-310 from CorrTest. EIS tests were performed with a frequency range of 1MHz to 1 kHz with a 5-mV voltage amplitude. By assuming that the solid polymer phase of the separator is nonconductive, the impedance measurements of the configured coin cells can be described by an equivalent circuit consisting of a capacity, an ionic resistance of the separator immersed in the electrolyte ($R_{ion}$), an electrical resistance of the coin cell's component ($R_{component}$) and an interfacial resistance between electrode and electrolyte ($R_{interfacial}$), which can be lumped into coin cell's resistance ($R_{cell}$). Therefore, the measured resistance of the coin cell can be expressed as

$$R_{cell} = R_{component} + R_{interfacial} + R_{ion} \tag{7}$$

### 3.2.3 Determination of the electrical tortuosity

Since the electrical conductivity of the bulk electrolyte is associated with the effective conductivity of the electrolyte in a porous medium, it is possible to derive indirect relationships between the electrical tortuosity and the effective conductivity of a porous medium. The average ionic resistance of a separator can be determined from multiple impedance measurements. The effective conductivity, $\sigma_{eff}$ (in S/m), of a separator filled with electrolyte, is

$$\sigma_{eff} = \frac{L_s}{R_{ion} A} \tag{8}$$

where the $R_{ion}$ (in Ω) is the ionic resistance of the separator saturated with electrolytes; $A$ (in m$^2$) is the electrode's Area, which is the spacer's area in this study; $L_s$ (in m) is the thickness of the separator in the coin cell. Similarly, the conductivity of the bulk electrolyte, $\sigma_0$ (in S/m), is given by

$$\sigma_0 = \frac{L}{R_0 A} \tag{9}$$

where the $R_0$ is the measured resistance of bulk electrolyte; $L$ is the distance between two electrodes, which is assumed to have the same distance as the thickness of the separator. To correlate the effective conductivity of the porous medium ($\sigma_{eff}$) with its porosity ($\Phi$) and electrical tortuosity ($\tau_e$), MacMullin number ($N_m$), which is defined as the ratio of $\sigma_0$ to $\sigma_{eff}$, was introduced in [14, 18], i.e.

$$N_m \equiv \frac{\sigma_0}{\sigma_{eff}} = \frac{\tau_e^2}{\Phi} \tag{10}$$

More details for the derived relationship of Eq.(10) can be found in [42]. From Eqs. (9) and (10), the ionic resistance of the separator immersed with electrolytes can be expressed as

$$R_{ion} = \frac{\tau_e^2 L_s}{\sigma_0 \Phi A} \quad (11)$$

where $\Phi$ and $\tau_e$ are the porosity and the electrical tortuosity of the separator, respectively. Therefore, by substituting Eq. (11) into Eq. (7), the overall resistance of the assembled coin cell, $R_{cell}$, can be expressed as [17]

$$R_{cell} = R_{series} + \frac{\tau_e^2 L_s}{\sigma_0 \Phi A} \quad (12)$$

$$R_{series} = R_{component} + R_{interfacial} \quad (13)$$

where $R_{cell}$ is the coin cell resistance interpreted from impedance test, and $R_{series}$ is the resistances of the series circuit representing the designed coin cell with zero thickness separator, including the resistances of the coin cell components (empty cell) (spring, two spacers, positive and negative cases) and the interfacial resistance between spacers and electrolytes when the resistance of the electrolytes is zero. $R_{series}$ can be determined from the designed coin cells by gradually reducing the separator layer number and extending the measured data to zero separator layer. Furthermore, $R_{component}$ can be obtained by measuring the resistance of the empty cell. Therefore, $R_{interfacia}$ can be determined from Eq. (13). Finally, the electrical tortuosity of the separator can be derived from Eq. (12).

### 3.3 Determination of the tortuosity using the 3D image-based method

#### 3.3.1 3D image-based reconstruction of the separator microstructure

The microstructure of the separator was reconstructed using 3D image data obtained from nanoscale XCT in Zernike Phase Contrast mode. The XCT scanning of the separator sample was performed in the Zeiss Xradia 810 Ultra CT system housed at Henry Moseley X-ray Imaging Facilities (HMXIF, The University of Manchester). A 5.4 keV quasi-monochromatic beam, an effective voxel size of 128.717 nm, and an exposure time of 30 s for each of 509 projections over 180 degrees were used. The raw image dataset was reorientated to ensure that the surface of the separator was perpendicular to the TTD of the separator. Then, the automated segmentation method, i.e., the ISODATA method embedded in the Avizo package, was applied to segment the polymer phase of the separator and distinguish it from the surrounding air phase. The detailed setup of the XCT scanning and segmentation method can be found in [24].

### 3.3.2 Geometrical tortuosity determined using the pore centroid method (PCM)

As introduced in Section 2.1, there are various methods to determine the geometrical tortuosity of the porous medium. PCM in commercial software Avizo was used in this study to determine the geometrical tortuosity of the porous separator due to its simplicity and straightforwardness. The PCM determines the geometrical tortuosity by calculating the average change of the pore centroid locations between the 2D slices in the 3D microstructures along a specified direction [34, 35, 43]. As shown in Fig. *3*, the path is formed by calculating the pore centroid coordinate $(x_i, y, z_i)$ on each plane and then linking these pore centroids. The geometrical tortuosity of the 3D microstructure can be determined as the ratio of the total path length ($L_g$) to the thickness of the sample ($L$)

$$\tau_g = \frac{L_g}{L} = \frac{\sum_{i=1}^{N-1} L_i}{L} \quad (14)$$

$$L_i = \sqrt{(x_{i+1} - x_i)^2 + (y_{i+1} - y_i)^2 + (z_{i+1} - z_i)^2} \quad (15)$$

where $i$ is the number of the 2D slice, and $N$ is the total number of the 2D slices in the microstructure.

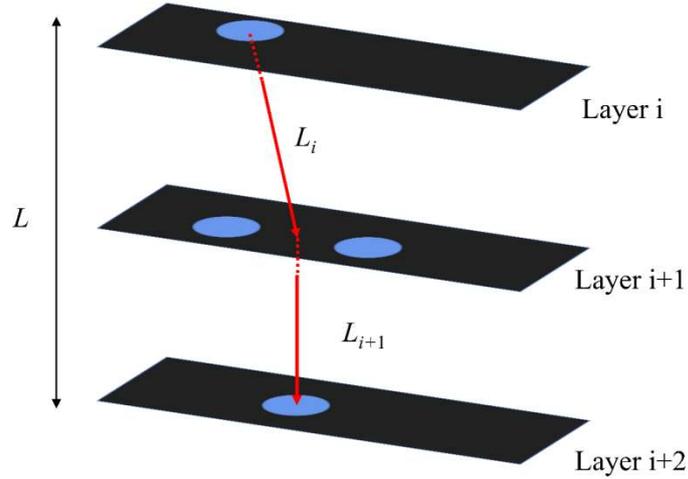

Fig. 3 The distances $L_i$ between the pore centroids on adjacent 2D slices

### 3.3.3 Electrical tortuosity determined using the 3D image-based method

The electrical tortuosity of the porous medium can be determined via various methods to simulate the electrical conduction in the porous medium, such as the finite difference method (FDM) [44], the random walk method (RWM) [45], the finite element method (FEM) [46], the lattice Boltzmann method (LBM) [47] and the finite volume method (FVM) [48]. In this work, the Avizo package of Xlab (i.e., the module of Formation Factor Experiment Simulation) was employed to simulate the transport of electric charges through the 3D microstructure of the separator. The solver of Avizo is based on the FVM [49] where the image voxels are directly used as the volume elements. The polymer phase of the separator is assumed to be homogeneous and insulating, and the pore phase of the separator is saturated with the electrolyte of electrical conductivity ($\sigma_0$). As shown in Fig. 4, a constant electrical potential difference is applied between the two opposite faces of the cubic microstructure of the separator (direct current is used). The other faces of the separator sample are enclosed with an electrical insulator. When the steady state is attained, the input and output current fluxes are equal. Using the charge conservation and Ohm's law on the entire volume, the effective conductivity ($\sigma_{eff}$ in S/m or A/V·m) can be calculated by

$$\frac{J_e^{pore}}{A} = \sigma_{eff} \frac{V_{in} - V_{out}}{L} \tag{16}$$

where $J_e^{pore}$ is the total electrical flux going through the input face of the porous separator sample (in A); $A$ is the area of the input face (in m²); $V_{in}$ and $V_{out}$ are respectively the imposed potentials at the input and output faces (in V); $L$ is the distance between the input and output faces (in m). The total electrical flux going through the input face can be calculated by locally applying Ohm's law

$$J_e^{pore} = \int_A -\sigma_0 \vec{\nabla} V \, dA \tag{17}$$

where $\vec{\nabla} V$ is the local gradient of the electrical potential (in V/m). Substituting Eq. (10) into Eq. (16), the electrical tortuosity, $\tau_e$, can be determined as

$$\tau_e = \sqrt{\frac{\sigma_0}{\sigma_{eff}}} \Phi = \sqrt{\frac{A\sigma_0 \frac{V_{in} - V_{out}}{L}}{J_e^{pore}}} \Phi \tag{18}$$

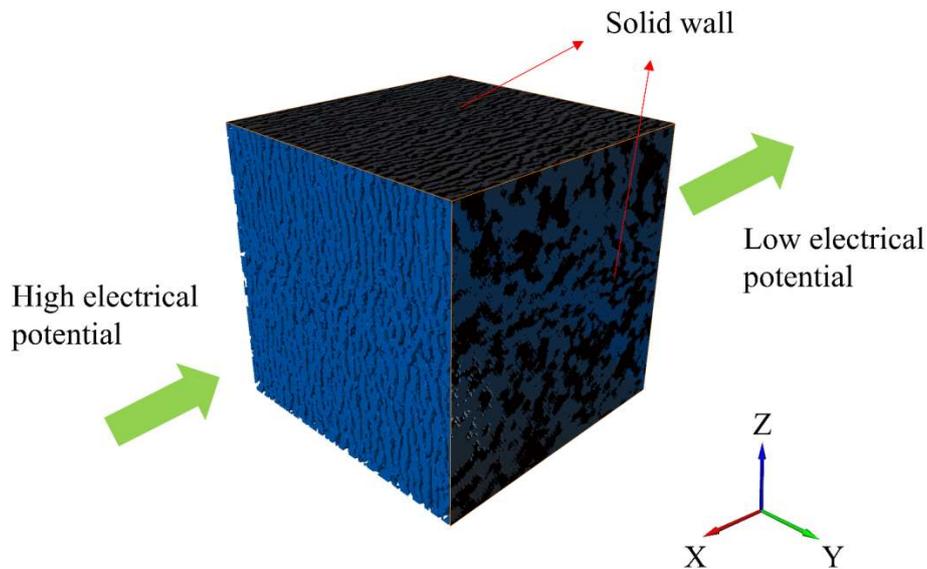

Fig. 4 Boundary conditions of the reconstructed separator for the electrical conduction simulation in Avizo along the X-axis direction (Blue phase stands for the pore of the separator).

## 4 Results and discussions

### 4.1 Empirical determination of the tortuosity

Based on the empirical Bruggeman equation, i.e., Eq. (3), the obtained tortuosity is 1.16 by inputting the porosity obtained from the manufacturer, which is 55% for Celgard 2500. It is important to understand the limitation of the Bruggeman equation, i.e., the microstructure difference is not considered in the Bruggeman equation when the porosities of the separators are the same. Practically, separators' tortuosities could be very different due to their different microstructures even though they have similar porosities. This limitation mainly comes from its original derivation, which was empirically determined by the experiments designed for spherical beans as the solid phase. However, the microstructure of the separator is not an ideal scenario. Therefore, it is indispensable to determine the tortuosity of a separator using the experimental or numerical method, which will be further discussed in the following sections.

## 4.2 Experimental determination of the electrical tortuosity based on the method developed in Section 3.2

To validate the porosity given by the supplier, the density method was employed to measure the porosity of the separator in this study. The obtained experimental results were listed in Table 2. Using Eq. (6), the porosity of the separator is calculated as 52% ± 0.9%, which is close to the nominal porosity (55%) provided by the supplier.

Table 2 The porosity measurement using the density method

|  | Weight in the air (g) | Weight in absolute ethanol (g) | Porosity (%) |
|---|---|---|---|
| Sample 1 | 0.00347 | 0.00056 | 52.9 |
| Sample 2 | 0.00344 | 0.00051 | 52.6 |
| Sample 3 | 0.00358 | 0.00056 | 51.1 |
| Sample 4 | 0.00355 | 0.00054 | 51.3 |

The ionic resistance of the separator filled with electrolyte was determined by EIS testing. It is necessary to ensure that sufficient electrolyte is added to wet the single-layer or stack-layered separators when the separators were sandwiched between the two spacers assembled in the coin cell. To achieve repeatable and consistent results, separators need to be fully wetted, for which the tester needs to wait for the separator to become transparent after adding the electrolyte. The impedance spectra of configurated coin cells and empty cells are respectively presented using Nyquist plots in Fig. *5* (a) and (b). The overall resistance of the coin cell $R_{cell}$ and the component resistance of the empty cell $R_{component}$ can be respectively interpreted from the intercepts of the Nyquist plots on the Re(Z) axis in Fig. *5* (a) and (b) [17]. From Fig. *5* (a), because of the increased path of length that ions move through the separators, the resistance shows that $R_{cell}$ shifts to higher resistance values with the larger thickness of separators. The component resistance $R_{component}$ can be directly interpreted from the Fig. 5 (b), which is $0.222 \pm 0.005$ Ω.

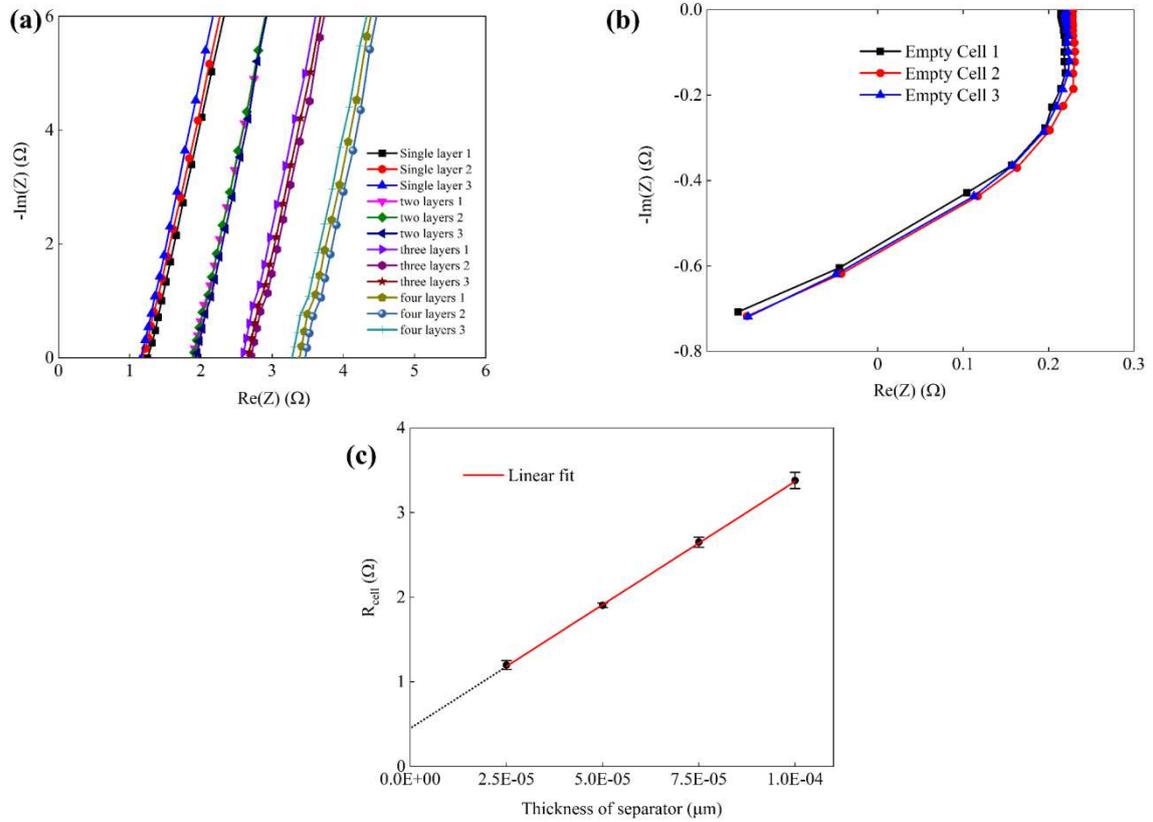

Fig. 5 Nyquist plots for (a) configured coin cells (spacer|separators|spacer) and (b) empty cells (spacer|spacer) (1,2,3 represent the repetitive tests); (c) The fitted curve of $R_{cell}$ with increased thickness of the separator.

As discussed in Section 3.2.3, to accurately determine the ionic conductivity of the separator immersed in the electrolyte, it is necessary to eliminate the possible error induced by series resistance. In Fig. 5 (c), the $R_{cell}$ is plotted as a function of the varied thickness of separators, which is ranged from 25 to 100 μm. According to Eq.(12), the fitted slope of the plot in Fig. 5 (c) is calculated as the $\frac{dR_{cell}}{dL_s} = \frac{\tau_e^2 L_s}{\sigma_0 \Phi A} = (2.913 \pm 0.04) \times 10^4$ Ω/m. By inputting the area of spacer ($A = 1.96 \times 10^{-4}$ $m^2$), the conductivity of bulk electrolytes ($\sigma_0 = 0.830$ $S/m$) taken from [40] and the experimentally measured porosity of the separator ($\Phi = 0.52$), the electrical tortuosity is determined as 1.57. Electrical tortuosity of the same separator was measured in [8] as 1.58 using the electrical conductivity experimental method, which is almost identical to the electrical tortuosity (1.57) measured in the present study. Furthermore, by extending the fitted line to intercept with the Y-axis shown in Fig. *5* (c), the series resistance of the configured coin cell $R_{series}$ is determined as 0.450 Ω. Therefore, the interfacial resistance between electrode and electrolyte in configured coin cell $R_{interfacial}$ can be derived as 0.228 Ω using Eq. (13).

### 4.3 Numerical determination of the geometrical and electrical tortuosity using the 3D image-based model

In this section, the 3D microstructure of the separator was firstly visualized using a 2D image dataset from nanoscale XCT scanning. The porosity was then analysed and compared with those given by the supplier. Subsequently, the geometrical and electrical tortuosities of the separator were determined numerically, as described briefly in Sections 3.3.2 and 3.3.3, which are detailed below. The module of the Pore Centroid Tortuosity in the Avizo was conducted to determine the geometrical tortuosity. The module of Formation Factor Experiment Simulation embedded in the XLab package of Avizo was adopted to calculate the effective electrical conductivity of the separator immersed in the electrolyte by the FVM described in Section 3.3.2, based on which the electrical tortuosity was determined by Eq. (10).

Before the 3D microstructure of the separator was used to determine tortuosity, it was used to analyse the porosity of the separator to validate the image-based model method by comparing the porosity provided by the supplier and the porosity data measured in Section 3.2.1. With the help of software in Avizo, the 3D microstructure of the separator was reconstructed in Fig. 6 (a). The macroscopic porosity measured by Avizo is 53.3%, which is comparable to the porosity of 55% from the supplier and the porosity of 52% measured using the density measurement method described in Section 3.2.1. Therefore, it can be reasonably inferred that the present nanoscale XCT can accurately determine the separator's porosity.

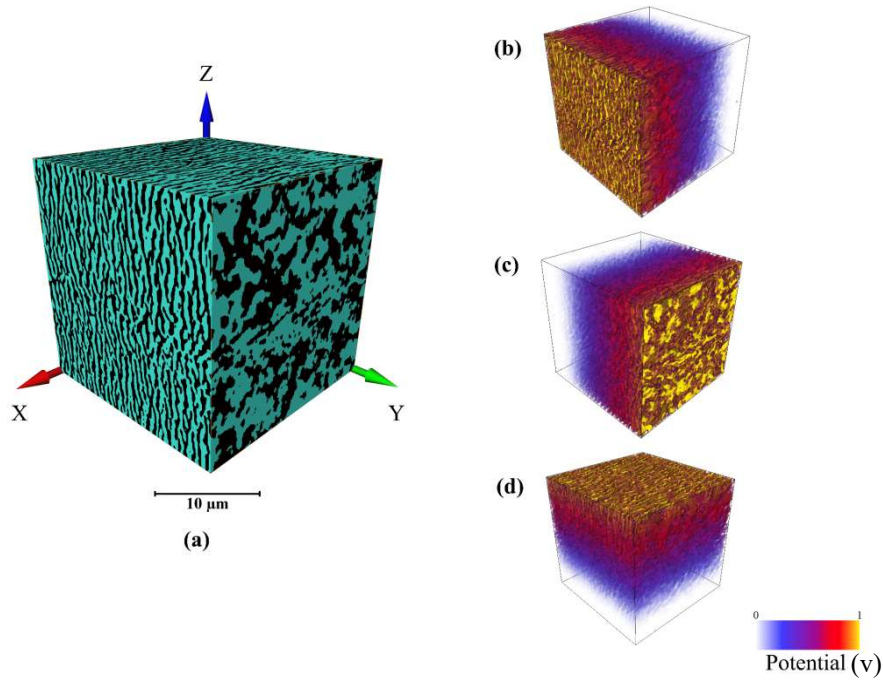

Fig. 6 (a) 3D microstructure of the separator reconstructed in Avizo (the pore phase and polymer phase are marked with green and black colour, respectively); (b), (c), (d) The electrical potential distributions inside the pore phase of the reconstructed separator along X, Y, Z directions, respectively.

Due to the intrinsic anisotropic properties of the Celgard 2500 separator, the tortuosities need to be analysed in three orthogonal directions, namely X, Y and Z. Based on the PCM, the geometrical tortuosities of the separator are respectively 1.06, 2.68 and 1.10. Furthermore, the electrical tortuosities are 1.20, 3.23 and 1.24, respectively, which are determined using a numerical method to simulate the electrical conductivity experiment via Avizo. The electrical potential distributions inside the pore phase of the separator are shown in Fig. 6 (b), (c) and (d). It can be observed that tortuosity along the Y direction (MD) is much higher than those in the other two directions, which is due to the uniaxial stretch of the separator experienced in MD. In addition, the reconstructed 3D microstructure of the porous separator can be used not only to calculate the geometrical tortuosity but also to design the microstructure of future separators. Lastly, based on the observation of the results for the geometrical and electrical tortuosity of the separator from numerical simulations, a more general relationship between the geometrical and physical tortuosities is proposed herein by introducing a phenomenological factor ($\beta$), i.e.

$$\tau_g = \beta\tau \tag{19}$$

where $\tau$ is the physical tortuosity, which can be related to various physical transport phenomena, including fluid permeation, electrical conduction, molecule diffusion and heat transfer. Therefore, based on the geometrical tortuosity ($\tau_g$) and electrical tortuosity ($\tau_e$) determined in Avizo, the electrical phenomenological factors $\beta_e$ in our study can be determined as 1.13, 1.21, and 1.13 along X, Y, and Z directions, respectively.

## 4.4 Comparisons of porosity, tortuosity and effective transport coefficient values obtained using different methods

In order to conveniently compare and quantify how the porosity and tortuosity affect the transport process of a porous medium, the effective transport coefficient ($\delta$), which is defined as the ratio of the porosity to the squared tortuosity (see the equation in Table 3) [8], is presented in this study. The effective transport coefficient is the inverse of the MacMullin number, which can be regarded as the normalization of the effective conductivity. Table 3 shows comparisons of porosity, tortuosity and effective transport coefficient ($\delta$) values determined using different methods.

Table 3 Comparisons of porosity and tortuosity for Celgard 2500 based on empirical, experimental and numerical methods (The number marked with * is the predicted value.)

| | Empirical method using Bruggeman equation | Experimental method using EIS experiment test | Numerical method using nanoscale XCT | |
|---|---|---|---|---|
| Porosity for the specified method (%) | $\Phi_{supplier} = 55.0$ (From supplier) | $\Phi_{exp} = 52.0$ (From density measurement) | $\Phi_{XCT} = 53.3$ | |
| Tortuosity | $\tau_{e\_Brugg} = 1.16$ | Electrical $\tau_{e\_Exp\_X} = 1.52^*$ $\tau_{e\_Exp\_Y} = 4.09^*$ $\tau_{e\_Exp\_Z} = 1.57$ | Geometrical $\tau_{g\_X} = 1.06$ $\tau_{g\_Y} = 2.68$ $\tau_{g\_Z} = 1.10$ | |
| Electrical phenomenological factors ($\beta_e$) | - | - | $\beta_{e\_XCT\_X} = 1.13$ $\beta_{e\_XCT\_Y} = 1.21$ $\beta_{e\_XCT\_Z} = 1.13$ | Electrical $\tau_{e\_XCT\_X} = 1.20$ $\tau_{e\_XCT\_Y} = 3.23$ $\tau_{e\_XCT\_Z} = 1.24$ |
| Effective transport coefficient ($\delta = \frac{\Phi}{\tau^2}$) | $\delta_{e\_Brugg} = 0.409$ | $\delta_{e\_Exp\_X} = 0.225^*$ $\delta_{e\_Exp\_Y} = 0.031^*$ $\delta_{e\_Exp\_Z} = 0.211$ | $\delta_{g\_X} = 0.474$ $\delta_{g\_Y} = 0.074$ $\delta_{g\_Z} = 0.440$ | |

It is evident that the geometrical tortuosity is the smallest, while the electrical tortuosity determined by the experimental method is smaller than the electrical tortuosity determined by the numerical method to simulate the electrical conduction. It can be explained that the very fine fibrils of the separator Celgard 2500 cannot be captured in the 3D microstructural model due to the limitation of the resolution of the nanoscale XCT, which also explains the observation that the experimentally determined porosity of the separator (52%) is less than the porosity measured from the nanoscale XCT method (53.3%).

In Table 3, the effective transport coefficient ($\delta$) along Z-direction derived from the EIS test is 0.211, which is approximately 52% of the effective transport coefficient obtained from the empirical Bruggeman equation ($\delta_{e\_Brugg} = 0.409$) and 61% of the effective transport coefficient from the electrical tortuosity using nanoscale XCT ($\delta_{e\_XCT\_Z} = 0.347$). In addition, due to the intrinsic thin thickness of the separator itself, it is difficult to experimentally determine the tortuosity of the separator along the in-plane directions (i.e., X and Y directions). However, the in-plane tortuosities of the separator can be determined using the numerical method to simulate the electrical conduction. Therefore, by assuming the value of the electrical tortuosity determined by the experimental method is proportional to that of electrical tortuosity determined by the numerical method along each specific direction (wherein $\tau_{e\_Exp\_Z} = 1.27\tau_{e\_XCT\_Z}$), the electrical tortuosities along the X and Y directions using the experimental method can be predicted as 1.52 and 4.09, respectively, when the same ratio value (1.27) along the Z direction is used. The effective diffusivity and conductivity of the electrolyte in a porous separator determined by both empirical equations and numerical simulation overestimate the performance of LIB in electrochemical modelling. By comparing with the empirical and numerical modelling methods, the EIS testing method is the most appropriate method to determine the electrical tortuosity of the separator.

## 5 Conclusions

In this paper, the tortuosities of separators, which plays an important role in the homogeneous modelling of the LIB, were determined based on the empirical Bruggeman equation, EIS experimental method and numerical simulation from 3D microstructure of nanoscale XCT images. The EIS experimental method is used to determine the electrical tortuosity of the separator and compared with the empirical Bruggeman equation and numerical modelling method, which gives a relationship of ($\tau_{e\_Brugg} < \tau_{e\_XCT} < \tau_{e\_Exp}$). Furthermore, a phenomenological factor ($\beta$) is proposed to correlate the relationship between geometrical tortuosity and physical tortuosity in a porous medium, whereby the electrical phenomenological factor is obtained based on the geometrical tortuosity and electrical tortuosity of the 3D microstructure of the separator using numerical simulation in Avizo software.

Although it is straightforward and convenient to determine the tortuosity of the separator using the empirical Bruggeman equation, it is not accurate to represent the tortuosity of separators with complex microstructure, especially when the two separators have similar porosity, but different microstructures.

The electrical tortuosity of the separator was determined using configured coin cells by EIS testing. The method proposed here can be standardised to measure the tortuosity of separators in LIB, and meanwhile, can also be used for thin and porous insulating membranes. In addition, the present study also showed that, in order to experimentally determine electrical tortuosity, the porosity of the separator also needs to be accurately measured rather than using the porosity provided by the manufacturing supplier. These porosity values have a relationship of ($\Phi_{exp} < \Phi_{XCT} < \Phi_{supplier}$).

The determination of the electrical tortuosity using microstructure obtained from XCT has the advantage to predict the tortuosity along the in-plane and through-plane directions while the experimental method can only measure its tortuosity along the through-plane direction. With the help of 3D microstructure, the geometrical tortuosity of the separator can be determined using PCM, which can support the design of future separators. In addition, the intrinsic anisotropic properties of the studied separator can be observed. Furthermore, by determining the proportional relationship between experimentally and XCT-determined electrical tortuosities along the TTD, the electrical tortuosity of the separator determined by the experimental method along X and Y directions can be predicted. However, compared with electrical tortuosity obtained from the EIS experimental method, the electrical tortuosity determined numerically based on the nanoscale XCT is generally smaller, i.e. ($\tau_{e\_XCT} < \tau_{e\_Exp}$), due to the limitation of the length resolution of the nanoscale XCT which cannot capture the fine fibrils of the separator in the XCT image and the FE model, and therefore, the separator model is less sinuous than the actual separator.

This study demonstrates the advantages and limitations of various methods to determine the tortuosity of the LIB's separator. By incorporating the experimental method to determine the tortuosity, the limitation of the nanoscale XCT can be compensated via manually adding fibres to reconstruct the microstructure of the separator before further conducting mechanical or electrochemical modelling.

**Acknowledgements**

This work was partially supported by the Henry Royce Institute for Advanced Materials, funded through EPSRC grants EP/R00661X/1, EP/S019367/1, EP/P025021/1 and EP/P025498/1. The authors acknowledge the Henry Mosley X-ray Imaging Facility, especially Dr Julia Behnsen for the technical support, advice and assistance provided. In addition, the authors also would like to acknowledge Mr Francis Moissinac's help with the experimental setup and suggestions.